\documentclass[a4paper,12pt]{article}
\usepackage{amssymb,amsthm,latexsym}
\newcommand{\Med}[1]{\left\langle #1 \right\rangle}
\newcommand{\med}[1]{\langle #1 \rangle}

\newtheorem{theorem}{Theorem}
\newtheorem{proposition}{Proposition}

\title{The high temperature region of the Viana-Bray diluted spin glass model\footnote{{\em 
Dedicated to Giovanni Jona Lasinio, in occasion of his 70th birthday}\newline}}

\author{
Francesco Guerra\footnote{e-mail:
{\tt francesco.guerra@roma1.infn.it}} \\
{\small {\itshape Dipartimento di Fisica, Universit\`a di Roma `La Sapienza'}}
\\
{\small {\itshape and INFN, Sezione di Roma, Piazzale A. Moro 2, 00185 Roma,
Italy}}\\
Fabio Lucio Toninelli\footnote{
e-mail: {\tt toninelli@eurandom.tue.nl}} \\
{\small {\itshape EURANDOM, P.O. Box 513 - 5600 Eindhoven, The Netherlands}}\\
{\small {\itshape and Institut f\"ur Mathematik der Universit\"at Z\"urich,}}\\
{\small {\itshape Winterthurer Strasse 190, CH-8057 Z\"urich, Switzerland
}}\
}

\date{\today}

\begin{document}

\maketitle

\begin{abstract}
In this paper, we study the high temperature or low connectivity phase of the
Viana-Bray model. This is a diluted version of the well known
Sherrington-Kirkpatrick mean field spin glass. In the whole
replica symmetric region, we obtain a complete control of the system,  proving
annealing for the
infinite volume free energy, and a central limit theorem for the
suitably rescaled fluctuations of the multi-overlaps. Moreover, we show that
free energy fluctuations, on the scale $1/N$, converge in the infinite
volume limit to a non-Gaussian random variable, whose variance diverges at the
boundary of the replica-symmetric region. The connection with the fully
connected Sherrington-Kirkpatrick model is discussed.
\end{abstract}

\newpage

\section{Introduction}

Diluted mean field spin
glasses attract a great interest among physicists and probabilists, for at least two reasons. First of
all, due to their finite degree of connectivity, they represent a sort of intermediate situation
between
fully connected models and realistic spin glasses with finite range interactions.
Secondly, many random optimization problems arising in theoretical computer
science are mapped in a natural way into the study of the ground state
of diluted mean field spin glass models. The mean field character of these systems makes them
exactly solvable, at least in the framework of Parisi theory of replica
symmetry breaking \cite{MPV}. Recently, many results have been obtained
in this direction, culminating in the resolution of the K-sat model
within the framework of ``one-step replica symmetry breaking'' in \cite{1RSB}.
Much less is know from the rigorous point of view, two remarkable
exceptions being Refs. \cite{franz-leone} and \cite{talaksat}. In
\cite{franz-leone}, through a suitable extension of the interpolation
methods introduced
in \cite{limterm} and \cite{broken} for fully connected models, S. Franz and
M. Leone proved, for a wide class of diluted models, that the thermodynamic
limit for the free energy density exists, and that it is bounded below by
Parisi solution with replica symmetry breaking. In Ref.
\cite{talaksat}, instead, M. Talagrand proved that replica symmetry holds for
sufficiently high temperature or low
average connectivity.

In the present work we concentrate on the case
of the Viana-Bray model \cite{viana}, \cite{kanter}, where each spin interacts
through two body couplings of random sign with a {\em finite} random number
of other spins, even in the infinite volume limit. This is a diluted version
of the well known Sherrington-Kirkpatrick (SK) model \cite{sk} \cite{MPV}.
We identify
the replica symmetric region, and we obtain a complete control of the system
there. In particular, through a suitable extension of the ``quadratic
replica coupling method'' we introduced in \cite{quadratic},  we prove that
annealing  holds for the free energy, in the infinite volume limit.
Moreover, as in \cite{CLT}, we prove limit theorems for the fluctuations of
(multi)-overlaps and of the free energy. While the fluctuations of the
multi-overlaps on the scale $1/\sqrt N$ turn out to be Gaussian in the
infinite volume limit, like for
the SK model, free energy fluctuations (on the scale $1/N$) tend to a
non-Gaussian random variable, whose variance diverges at the boundary
of the replica symmetric region.

The organization of the paper is as follows. In Section 2 we give the basic
definitions concerning the model, and in Section 3 we discuss
the role played by the multi-overlaps in its thermodynamical description.
The relationship
between the model under consideration and the fully connected one is
considered in Section 4. In Sections 5 and 6, we identify the replica symmetric
region and we prove annealing for the free energy. Finally, in Sections
7 and 8 we prove limit theorems for the fluctuations in the annealed region, while
Section 9 is dedicated to conclusions and outlook to future developments.

\section{Definition of the model}

The Hamiltonian of the Viana-Bray model \cite{viana},
for a given configuration of the
$N$ Ising spin variables $\sigma_i=\pm1, i=1,\ldots,N$, is defined as
\begin{equation}
\label{H}
H_N(\sigma,\alpha;{\cal J})=-\sum_{\mu=1}^{\xi_{\alpha N}}
J_\mu \sigma_{i_\mu} \sigma_{j_\mu}.
\end{equation}
Here, $\xi_{\alpha N}$ is a Poisson random variable of mean value $\alpha N$,
for some $\alpha>0$, {\em i.e.},
\begin{equation}
P(\xi_{\alpha N}=k)=\pi(k,\alpha N)\equiv
e^{-\alpha N}\frac{(\alpha N)^k}{k!}\;\;\;k=0,1,2,\ldots,
\end{equation}
while $\{J_\mu\}$ is a family of independent identically distributed (i.i.d.)
symmetric random variables, and $i_\mu,j_\mu$ are i.i.d. integer valued
random variables, uniformly distributed on the set $\{1,2,\ldots,N\}$.
We denote by ${\mathcal J}$ the dependence of the Hamiltonian on the whole
set of quenched disordered variables $\xi_{\alpha N}, J_\mu, i_\mu,j_\mu$.
The parameter $\alpha$ fixes the average degree of connectivity of the system. Indeed
the number of different sites, which interact with a given spin variable, behaves
approximately like a Poisson random variable of parameter $2\alpha$, for large
values of $N$. This is to be compared with the case of the SK
model, where any spin interacts with all the other $N-1$.
A second important difference with respect to  the SK model is that,
in the present case, the infinite volume limit of the system {\em does}
depend on the probability distribution $\rho(J)$ of
$J_\mu$. In the case $\rho(J)=1/2(\delta(J-1)+\delta(J+1))$,
the Viana-Bray model is closely related to the so called 2-XOR-SAT
problem \cite{xor-sat} of computer science.
In the course of this work, we do not specify the form of $\rho(J)$, but for simplicity we
assume $J$ to be a bounded random variable
\begin{equation}
\label{jbounded}
|J|\le 1.
\end{equation}
More general cases can be considered, at the expense of some additional
technical work.

The partition function $Z_N(\beta,\alpha; {\cal J})$, the disorder dependent
free energy $f_N(\beta,\alpha; {\cal J})$, the Gibbs
state $\omega_{{\cal J}}$ and
the quenched free energy $-\beta A_N(\beta,\alpha)$ are defined in the
usual way, for a given value of the inverse temperature $\beta$:
\begin{eqnarray}
&&\label{Z}
Z_N(\beta,\alpha; {\cal J})=
\sum_{\{\sigma\}} e^{-\beta H_N(\sigma,\alpha; {\cal J})}\\
&&f_N(\beta,\alpha; {\cal J})=-\frac1{N\beta}\ln Z_N(\beta,\alpha; {\cal J})\\
&&\omega_{\cal J}({\cal O})=Z_N(\beta,\alpha; {\cal J})^{-1}
\sum_{\{\sigma\}} {\cal O}(\sigma)e^{-\beta H_N(\sigma,\alpha; {\cal J})}\\
\label{AN}
&& A_N(\beta,\alpha)=\frac1NE\ln Z_N(\beta,\alpha; {\cal J})=-\beta E
f_N(\beta,\alpha; {\cal J}).
\end{eqnarray}
Here, ${\mathcal O}$ is a generic function of the spin variables,
and $E$ denotes expectation with
respect to all quenched random variables:
\begin{equation}
E(.)=E_{\xi_{\alpha N}}E_{\{J_\mu\}}E_{\{i_\mu\}}E_{\{j_\mu\}}(.).
\end{equation}
Like in the case of fully connected models, it is possible to prove
that $f_N(\beta,\alpha; {\cal J})$ is self-averaging
when the system size grows to infinity, and to give bounds,
exponentially small in $N$, on the probability of its fluctuations.
The precise result is stated and proved in Appendix A.

As usual, one introduces {\em real} replicas as independent identical copies
of the system, subject to the same disorder realization, and denotes with
$\Omega_{\cal J}(.)$ the disorder dependent
product Gibbs state
\begin{equation}
\Omega_{\cal J}=\omega_{\cal J}^{(1)}\otimes\omega_{\cal J}^{(2)}\otimes
\ldots,
\end{equation}
where the state $\omega_{\cal J}^{(a)}$ acts on the
$a$-th replica.
Moreover, the average $\med{.}$, involving both thermal and disorder averages,
is defined as
\begin{equation}
\med{.}=E\,\Omega_{\cal J}(.).
\end{equation}

A very important role is played by the multi-overlaps between
$n$ configurations $\sigma^{(1)},\ldots,\sigma^{(n)}$, defined as
\begin{equation}
q_{1\ldots n}=\frac1N\sum_{i=1}^N\sigma^{(1)}_i\ldots\sigma^{(n)}_i.
\end{equation}
Of course,
\begin{equation}
-1\le q_{1\ldots n}\le 1.
\end{equation}
Notice that for $n=2$ one recovers the usual definition of the overlap
as normalized scalar product between two configurations.

While for fully connected models the whole physical content of the
theory is encoded in the probability distribution of the overlaps \cite{MPV},
all multi-overlaps play an essential role in the present case \cite{viana}, \cite{kanter}.
In Section 4 we will show
how, when the limit of infinite connectivity is suitably performed, the
multi-overlaps  with $n>2$ become inessential.

\section{The role of the multi-overlaps}

An important ingredient of the methods employed in \cite{franz-leone} is
a smart use of the properties of the Poisson random variables.
Indeed, while the choice of the Poisson distribution for the
number $\xi_{\alpha N}$ of terms appearing in the Hamiltonian
(\ref{H}) is in principle not essential (any other random variable sharply
concentrated around the value $\alpha N$ would yield an equivalent model,
in the infinite volume limit), it turns out to be a great technical
simplification.
The basic elementary properties one employs, for the distribution function
of a Poisson random variable $\xi_\lambda$ of parameter $\lambda>0$,  are
\begin{equation}
\label{prop1}
k\,\pi(k,\lambda)=\lambda\,\pi(k-1,\lambda)
\end{equation}
and
\begin{equation}
\label{prop2}
\frac d{d\lambda}\pi(k,\lambda)=-\pi(k,\lambda)+\pi(k-1,\lambda)(1-\delta_{k,0}).
\end{equation}
In a sense, Eq. (\ref{prop1}) replaces the identity
\begin{equation}
E J F(J)=E F'(J),
\end{equation}
which plays a fundamental role in the study of the fully connected models,
and which holds for any smooth function $F$ if $J$ is a Gaussian standard
random variable.

For instance, let us show how Eq.  (\ref{prop1})  allows to
express the internal energy of the Viana-Bray model as a sum of
simple averages involving multi-overlaps. For an analogous computation, see Ref. \cite{franz-leone}. One has
\begin{eqnarray}
-\frac{\partial}{\partial\beta}A_N(\beta,\alpha)=\frac{\med{H}}N=-\frac1N
\sum_{k=1}^\infty \pi(k,\alpha N)\sum_{\mu=1}^k\Med{J_\mu \sigma_{i_\mu}
\sigma_{j_\mu}}_{k},
\end{eqnarray}
where $\med{.}_{k}$ denotes the average where the value of
the random variable $\xi_{\alpha N}$ has been fixed to $k$.
Then, using property (\ref{prop1}),
\begin{equation}
\label{then}
\frac{\med{H}}N=-\frac1N\sum_{k=1}^\infty k \,\pi(k,\alpha N)\Med{J_k\sigma_{i_k}
\sigma_{j_k}}_{k}=
-\alpha\sum_{k=1}^\infty \pi(k-1,\alpha N)\Med{J_k\sigma_{i_k}
\sigma_{j_k}}_{k}.
\end{equation}
Now, we  use the identity
\begin{eqnarray}
\Med{J_k\sigma_{i_k}\sigma_{j_k}}_{k}=
E\,\omega_{\cal J}(J_k\sigma_{i_k}\sigma_{j_k})_{k}=
E\frac{\omega_{\cal J}\left(J_k\sigma_{i_k}\sigma_{j_k}
\exp(\beta J_k\sigma_{i_k}\sigma_{j_k})\right)_{k-1}}
{\omega_{\cal J}(\exp(\beta J_k\sigma_{i_k}\sigma_{j_k}))_{k-1}},
\end{eqnarray}
to rewrite (\ref{then}) as
\begin{eqnarray}
&&\frac{\med{H}}N=-\alpha E \frac{\sum_{i,j=1}^N}{N^2}\frac{
\omega_{\cal J}(J\sigma_{i}\sigma_{j}
\exp(\beta J\sigma_{i}\sigma_{j}))}{\omega_{\cal J}
(\exp(\beta J\sigma_{i}\sigma_{j}))}\\
&&
\label{now}
=-\frac{\alpha}{N^2}\sum_{i,j=1}^N E J
\frac{\tanh(\beta J)+\omega_{\cal J}(\sigma_i
\sigma_j)}{1+\tanh(\beta J)\omega_{\cal J}(\sigma_i
\sigma_j)},
\end{eqnarray}
where $J$ is independent of the couplings $J_\mu$ on which $\omega_{\cal J}$ depends.
Notice that we have employed the identity
\begin{equation}
e^{\beta J\sigma_i\sigma_j}=\cosh(\beta J)+\sigma_i\sigma_j\sinh(\beta J)
\end{equation}
in the last step.
Thanks to (\ref{jbounded}), $|\tanh(\beta J)|\le\tanh\beta<1$ so that the
expression in (\ref{now}) can be expanded in absolutely convergent
Taylor series around $\tanh(\beta J)=0$. Recalling the definition of
the multi-overlaps and the symmetry of the random variable $J$, one finally
finds
\begin{eqnarray}
-\frac{\partial}{\partial\beta}A_N(\beta,\alpha)=\frac{\med{H}}N&=&
-\alpha E(J\tanh(\beta J))\\\nonumber
&&+\alpha
\sum_{n=0}^\infty \med{q^2_{1\ldots 2n+2}}\,
E\left\{J\tanh^{2n+1}(\beta J)(1-\tanh^2(\beta J))\right\}.
\end{eqnarray}
In the particular case where $J_\mu=\pm 1$, the above expression reduces to
\begin{equation}
-\frac{\partial}{\partial\beta}A_N(\beta,\alpha)=-\alpha \tanh \beta
+\alpha
\sum_{n=0}^\infty (\tanh\beta)^{2n+1}(1-\tanh^2\beta)
\med{q^2_{1\ldots 2n+2}}.
\end{equation}

\section{The infinite connectivity limit and the SK model}

In this section, we discuss the relationship between the Viana-Bray and the
fully connected SK model. As it was already observed in
\cite{viana} \cite{kanter}, the latter is obtained when the average connectivity $\alpha$
tends to infinity, provided that the strength of the couplings $J_\mu$,
or equivalently the inverse temperature, is suitably rescaled to zero.
Let us discuss this point in greater detail. To this purpose, recall that the
 SK model in zero external field
is defined by the Hamiltonian
\begin{equation}
\label{Hsk}
H^{S.K.}_N(\sigma;J)=
-\frac1{\sqrt N}\sum_{1\le i<j\le N}J_{ij}\sigma_i\sigma_j,
\end{equation}
where the couplings $J_{ij}$ are i.i.d. centered Gaussian random
variables of unit variance. Now, we want to compare the Viana-Bray model, with parameters
$\beta$ and $\alpha$, with the SK model, at an inverse
temperature $\beta'$ defined as
\begin{equation}
\label{beta'}
\beta'^2=2\alpha E\tanh^2(\beta J).
\end{equation}
In particular we show that, in the limit $\alpha\to\infty$, $\beta\to 0$
with $\beta'=const$, one has
\begin{equation}
\label{rapporto}
\left\{\lim_{N\to\infty}\frac1NE\ln Z_N(\beta,\alpha;{\cal J})
\right\}\stackrel{\alpha\to\infty}{\longrightarrow}   \left\{
\lim_{N\to\infty}\frac1NE\ln Z^{S.K.}_N(\beta';{J})\right\}.
\end{equation}
To this purpose, let $0\le t\le1$ and define an auxiliary partition function
$Z_N(t)$ as
\begin{equation}
\label{Zt}
Z_N(t)=\sum_{\{\sigma\}}\exp\left(\beta\sum_{\mu=1}^{\xi_{\alpha N t}}
J_\mu \sigma_{i_\mu}\sigma_{j_\mu}+\beta'\sqrt{\frac{1-t}N}\sum_{1\le i<j\le N}
J_{ij}\sigma_i\sigma_j\right).
\end{equation}
Of course, for $t=1$ one recovers the partition function (\ref{Z}) of the
diluted model, while for $t=0$ one has the partition function of the fully
connected model, at inverse temperature $\beta'$.
The $t$ derivative of $1/NE\ln Z_N(t)$ can be performed along the lines of the
computation of $\partial_\beta A_N(\beta, \alpha)$ in the previous section, with the result
\begin{eqnarray}
\label{a1}
\frac{d}{dt}\frac1NE\ln Z_N(t)&=&\alpha\left(E \ln \cosh (\beta J)-\frac1
2\sum_{n=1}^\infty\frac{E\tanh^{2n}(\beta J)}n
\med{q_{1\ldots 2n}^2}\right)\\
\label{a2}
&&-\frac{\beta'^2}4\left(1-\med{q_{12}^2}\right)\\
\nonumber
&=&\left(\alpha E\ln \cosh(\beta J)-\frac{\beta'^2}4\right)
-\frac\alpha 2\sum_{n=2}^\infty\frac{E(\tanh^{2n}(\beta J))}n
\med{q_{1\ldots 2n}^2}.
\end{eqnarray}
The term (\ref{a1}) derives from the $t$ dependence of the Poisson
random variable $\xi_{\alpha N t}$ in (\ref{Zt}), while (\ref{a2})
comes from the $\sqrt{1-t}$ factor which multiplies the SK Hamiltonian.
Before we proceed, let us notice that we have proved the inequality, uniform in $N$,
\begin{equation}
\label{uniform}
\frac d{dt}\frac1NE\ln Z_N(t)\le \alpha E\ln \cosh(\beta J)-\frac{\beta'^2}4,
\end{equation}
whose implications will be discussed below. Now, it is easy to see that
the $t$ derivative we are considering vanishes for $\beta\to0$, $\alpha\to\infty$, if the
constraint (\ref{beta'}) is satisfied. Indeed, for $\alpha\to\infty$
Eq. (\ref{beta'}) reduces to
\begin{equation}
2\alpha \beta^2E J^2=\beta'^2+O\left(\frac1\alpha\right),
\end{equation}
so that
\begin{equation}
\alpha E\ln \cosh(\beta J)-\frac{\beta'^2}4=
\alpha E\ln (1+\frac{\beta^2J^2}2)-\frac{\beta'^2}4+
O\left(\frac1\alpha\right)=O\left(\frac1\alpha\right)
\end{equation}
and
\begin{eqnarray}
\frac\alpha2\sum_{n=2}^\infty \frac{E\tanh^{2n}(\beta J)}n
\med{q_{1\ldots 2n}^2}\le\frac\alpha2\sum_{n=2}^\infty\frac{\beta^{2n}
}n\stackrel{\alpha\to\infty}{\longrightarrow} 0,
\end{eqnarray}
which concludes the proof of (\ref{rapporto}).
\hfill $\Box$

\section{The replica symmetric bound and the annealed region}

In \cite{broken} it was proven that the Parisi solution for the
SK model, with an arbitrary number of levels of replica symmetry
breaking, is a lower bound for the free energy, at any temperature.
Along the same lines, this result was extended in \cite{franz-leone}
to the case of diluted models. In this context, one has to face the
additional difficulty that, even at the level of the replica symmetric
approximation, the Parisi order parameter is a function \cite{kanter} (the
probability distribution of the effective field) rather than a single
number, as it happens instead for fully connected models \cite{MPV}.
In the present section, we recall briefly the replica symmetric bound
for the Viana-Bray model under consideration,
and we discuss the high temperature or low connectivity phase, where
this bound actually gives the correct limit.

Let $g$ be an arbitrary symmetric random variable (we assume its
distribution to be regular enough to guarantee that all expressions below
are well defined), and define the random variable $u$ as
\begin{equation}
  \tanh(\beta u)=\tanh(\beta J)\tanh (\beta g).
\end{equation}
Here, $J$ is distributed like any of the couplings $J_\mu$ and is
independent of them (as well as of $g$). For given $\beta$ and $\alpha$,
the replica symmetric trial functional $F_{RS}(\beta,\alpha;g)$ is defined as
\begin{eqnarray}
\nonumber
F_{RS}(\beta,\alpha;g)&=&\ln 2+\alpha E \ln \cosh (\beta J)+E\ln
\cosh\bigl(\beta\sum_{\ell=1}^{\xi_{2\alpha}}u_\ell\bigr)-
2\alpha E\ln\cosh (\beta u)
\\
&&-\frac\alpha2E \ln \left
(1-\tanh^2(\beta J)\tanh^2(\beta g_1)\tanh^2(\beta g_2)\right).
\end{eqnarray}
Here, $u_\ell$ are independent copies of $u$ and $g_1,g_2$ are
independent
copies of $g$. Then, one has \cite{franz-leone}
\begin{equation}
\label{boundRS}
  \frac1N E\ln Z_N(\beta,\alpha;{\cal J})\le \inf_g
  F_{RS}(\beta,\alpha;g)+O\left(\frac1N
\right),
\end{equation}
where the infimum is taken over the space of symmetric random
variables $g$. It is not difficult to see, computing the functional derivative
of $F_{RS}(\beta,\alpha;g)$ with respect to the probability distribution
$P(g)$ of $g$, that a sufficient condition of
extremality for the replica symmetric functional is \cite{kanter}
\begin{equation}
\label{self-cons}
  g\stackrel{d}{=}\sum_{\ell=1}^{\xi_{2\alpha}}u_\ell=
\frac1\beta
\sum_{\ell=1}^{\xi_{2\alpha}}\tanh^{-1}(\tanh(\beta J_\ell)\tanh
(\beta g_\ell)),
\end{equation}
where the equality holds in distribution. It is clear that the above
equation always admits the trivial solution $g$ concentrated
at the value zero, {\em i.e.}, with
$P(g)=\delta(g)$. In this case,
\begin{eqnarray}
F_{RS}(\beta,\alpha;g\equiv 0)=\ln 2+\alpha E\ln \cosh (\beta J)
\end{eqnarray}
which corresponds to take the expectation with respect to the
the random coupling signs before the logarithm, in the definition (\ref{AN}) of $A_N(\beta,\alpha)$:
\begin{eqnarray}
F_{RS}(\beta,\alpha; 0)=\frac1N E\ln E_{\{sign(J_\mu)\}}Z_N(\beta,\alpha;{\cal J}).
\end{eqnarray}
In the following, we call $-1/\beta F_{RS}(\beta,\alpha;0)$ the ``annealed free energy".
The following result shows
that, in a certain region of the parameters $\beta$ and $\alpha$, the
trivial solution of (\ref{self-cons}) is actually the only one:
\begin{proposition}
 \label{prop:contraz}
If
\begin{equation}
\label{ann_reg}
2\alpha E\tanh^2(\beta J)<1\;\;\;(annealed\; region),
\end{equation}
the only symmetric random variable $g$ satisfying equation
(\ref{self-cons}) is the degenerate one: $P(g)=\delta(g)$.
\end{proposition}
Notice that, for $\alpha<1/2$, the annealed region extends up to $\beta=\infty$.

{\em Proof of Proposition \ref{prop:contraz}}.
Let
\begin{equation}
  \phi(v)=E\, e^{i v g}
\end{equation}
be the characteristic function of $g$,  which can be rewritten, thanks
to condition (\ref{self-cons}), as
\begin{equation}
\label{primopasso}
  \ln \phi(v)= 2\alpha \left(E\, \exp\left(i\frac v\beta \tanh^{-1}(\tanh(\beta
J)\tanh(\beta g))\right)-1\right).
\end{equation}
This implies that
\begin{equation}
|\ln \phi(v)|\le 2\alpha |v|\sqrt{2\alpha E\tanh^2(\beta J)},
\end{equation}
where we used the fact that
$$
E g^2\le 2\alpha,
$$
as it easily follows from (\ref{self-cons}) and from
$|J|\le1$.
Now, one can iterate the procedure, replacing the random variable $g$ which appears at the
right hand side of (\ref{primopasso}) with the expression given by
Eq. (\ref{self-cons}), and so on. At the $n$-th step of the iteration
one has the bound
\begin{equation}
\left|\ln \phi(v)\right|\le 2\alpha |v|
\left(2\alpha E\tanh^2(\beta J)\right)^{n/2}.
\end{equation}
which goes to zero when $n\to\infty$,
if condition (\ref{ann_reg}) holds. \hfill $\Box$

On the other hand it is easy to realize that, outside the annealed
region, the choice of the identically vanishing $g$ does not
realize the infimum in (\ref{boundRS}).
Indeed, consider even the simple case of a two-valued random variable
$g$  with distribution
$$
P(g)=\frac12(\delta(g-g_0)+\delta(g+g_0)).
$$
When $g_0\simeq 0$, one finds
\begin{eqnarray}
F_{RS}(\beta,\alpha;g)-F_{RS}(\beta,\alpha;g\equiv0)=\frac\alpha2\beta^4
g_0^4(1-2\alpha E\tanh^2(\beta J))+O(g_0^6),
\end{eqnarray}
which is negative if (\ref{ann_reg}) does not hold.

It is interesting to observe that breaking of annealing
outside the region (\ref{ann_reg}) can also be proved through a comparison
with the SK model. Indeed, integration of the inequality
(\ref{uniform}) with respect to $t$ between 0 and 1 gives
\begin{eqnarray}
\nonumber
\ln 2+\alpha E\ln \cosh(\beta J)-\frac1NE\ln Z_N(\beta,\alpha;{\cal J})
\ge \ln 2+\frac{\beta'^2}4-\frac1NE\ln Z_N^{S.K.}(\beta';J),
\end{eqnarray}
{\em i.e.}, the difference between the quenched and the annealed free energies
is larger (in absolute value) for the diluted model than for its
fully connected counterpart if
$\beta,\alpha$ and $\beta'$ are related by the condition (\ref{beta'}).
Therefore, since it is well known that
\begin{equation}
\lim_{N\to\infty}\frac1NE\ln Z_N^{S.K.}(\beta';J)<\ln 2+\frac{\beta'^2}4
\end{equation}
for $\beta'>1$, one has immediately
breakdown of annealing for the Viana-Bray model,
when $\beta'^2=2\alpha E\tanh^2(\beta J)>1$.


\section{Control of the annealed region}

\label{sez_ann}

In the present section we prove that annealing actually holds for the
Viana-Bray model in the region of parameters (\ref{ann_reg}), {\em i.e.},
that
\begin{theorem}
\label{teoannealing}
For $2\alpha E\tanh^2(\beta J)<1$,
\begin{eqnarray}
\frac1N E\ln Z_N(\beta,\alpha;{\cal J})= \ln 2+\alpha E\ln \cosh(\beta J)
+O\left(\frac1N\right).
\end{eqnarray}
\end{theorem}
We prove the theorem via a suitable adaptation of the ``quadratic replica
coupling'' method we introduced in \cite{quadratic} for the SK model.
While the above result can also be obtained through the ``second moment
method'' \cite{talagcorso}, which consists in showing that
\begin{equation}
\frac1N \ln E (Z_N)^2=\frac 1N\ln (E Z_N)^2+o(1),
\end{equation}
the quadratic method we employ allows us to obtain self-averaging of the multi-overlaps in a stronger form,
and to prove limit theorems for the fluctuations, as shown in the next two sections.

Consider a system of two coupled replicas of the model,
defined by the partition function
\begin{eqnarray}
Z_N^{(2)}(\beta,\alpha,\lambda;{\cal J})=\sum_{\{\sigma^1,\sigma^2\}}
e^{-\beta H_N(\sigma^1,\alpha;{\cal J})-
\beta H_N(\sigma^2,\alpha;{\cal J})+N\frac{\lambda}2
\, q_{12}^2,
}
\end{eqnarray}
where $\lambda\ge 0$. Notice that the quadratic interaction gives a large
weight to the pairs of  configurations whose overlap is different from zero.
Like in \cite{quadratic} the idea is to show that, if $\lambda$ is not
too large, the interaction does not modify the infinite
volume free energy density, so that $q_{12}$ must be typically close to zero.
Indeed, we can prove
\begin{theorem}
\label{teoaccopp}
In the region
\begin{eqnarray}
\label{triangular}
(\lambda+2\alpha E\tanh^2(\beta J))<1,\hspace{.5cm}
\lambda,\alpha\ge0,
\end{eqnarray}
one has
\begin{eqnarray}
\label{stat1}
\frac 1{2N}E\ln Z_N^{(2)}(\beta,\alpha,\lambda;{\cal J})=
\ln 2+\alpha E\ln \cosh (\beta J)+O\left(\frac1N\right)
\end{eqnarray}
and
\begin{eqnarray}
\label{stat2}
\med{q_{1\ldots 2n}^2}\le\med{q_{12}^2}=O\left(\frac1N\right).
\end{eqnarray}
\end{theorem}
Of course, this result implies the previous Theorem 2 since,  for $\lambda=0$,
$$
\frac{1}{2N} E\ln Z_N^{(2)}(\beta,\alpha,0;{\cal J})=\frac1N E\ln Z_N
(\beta,\alpha;{\cal J}).
$$
{\em Proof of Theorem \ref{teoaccopp}}. First of all, since \cite{franz-leone}
\begin{eqnarray}
\frac \partial{\partial\alpha} \frac1N E\ln Z_N(\beta,\alpha;{\cal J})=
E\ln \cosh(\beta J)-\frac12\sum_{n=1}^\infty\frac{E\tanh^{2n}(\beta J)}n
\med{q^2_{1\ldots 2n}},
\end{eqnarray}
and
\begin{equation}
\med{q^2_{1\ldots 2n}}=\frac1{N^2}\sum_{i,j=1}^N E \omega^{2n}_{\cal J}(
\sigma_i\sigma_j)\le \frac1{N^2}\sum_{i,j=1}^N E \omega^{2}_{\cal J}(
\sigma_i\sigma_j)=\med{q^2_{12}},
\end{equation}
one can write
\begin{eqnarray}
\frac \partial{\partial\alpha} \left(F_{RS}(\beta,\alpha;0)-
\frac1NE\ln Z_N\right)\le \frac {\med{q_{12}^2}}2
E \ln(1-
\tanh^2(\beta J))^{-1}.
\end{eqnarray}
Therefore, using convexity of $\ln Z_N^{(2)}$ with respect to $\lambda$
and the identity
\begin{equation}
\left.\frac{\partial}{\partial \lambda}
\frac 1{2N}E\ln Z_N^{(2)}(\beta,\alpha,\lambda;{\cal J})\right|_{\lambda=0}
=\frac14 \med{q_{12}^2},
\end{equation}
one has
\begin{eqnarray}
\label{sothat}
\frac \partial{\partial\alpha} \left(F_{RS}(\beta,\alpha;0)-\frac
{E\ln Z_N}N\right)\le \frac{2E \ln(1-\tanh^2(\beta J))^{-1}}{\lambda
}\left(\frac{E\ln Z_N^{(2)}(\lambda)}{2N}-
\frac{E\ln Z_N}N\right).
\end{eqnarray}
Next, we need an upper bound for $1/(2N) E\ln Z_N^{(2)}$. To this purpose,
we take $\lambda$ to depend on $\alpha$ as $\lambda(\alpha)=\lambda_0-
2\alpha E\tanh^2(\beta J)$, and we compute
\begin{eqnarray}
&&\frac{d}{d\alpha}\frac1{2N}E\ln Z_N^{(2)}(\beta,\alpha,\lambda(\alpha);
{\cal J})=-\frac12E\tanh^2(\beta J)\med{q_{12}^2}_{\alpha,\lambda(\alpha)}+
E\ln \cosh(\beta J)\\\nonumber
&&+\frac1{4N^2}\sum_{i,j=1}^NE\ln \left[
(1+\tanh^2(\beta J)\Omega_{\alpha,\lambda(\alpha)}(\sigma^1_i
\sigma^1_j\sigma^2_i\sigma^2_j))^2-4 \tanh^2(\beta J)
\omega^2_{\alpha,\lambda(\alpha)}(\sigma_i
\sigma_j)\right],
\end{eqnarray}
where we employed Eq. (\ref{prop2}) and the symmetry of $J$.
Here, the averages refer to the coupled
system with parameters $\alpha,\lambda(\alpha)$. Then,
\begin{eqnarray}
\frac{d}{d\alpha}\frac1{2N}E\ln Z_N^{(2)}(\beta,\alpha,\lambda(\alpha);
{\cal J})&\le& -\frac12E\tanh^2(\beta J)\med{q_{12}^2}_{\alpha,
\lambda(\alpha)}\\\nonumber
&&+
E\ln \cosh(\beta J)+\frac12 \ln (1+E\tanh^2(\beta J)\med{q_{12}^2})\\
&\le& E\ln \cosh(\beta J),
\end{eqnarray}
where we used Jensen's inequality to take expectation inside the logarithm,
and the elementary estimate
$$\ln (1+x)\le x.
$$
Therefore, integrating between 0 and $\alpha$ one has
\begin{eqnarray}
\label{integrando}
\frac1{2N}E\ln Z_N^{(2)}(\beta,\alpha,\lambda;{\cal J})&\le&
\alpha E\ln\cosh(\beta J)+\frac1{2N}\ln \sum_{\{\sigma^1,\sigma^2\}}
e^{N {\lambda_0}\,q_{12}^2/2},
\end{eqnarray}
since at $\alpha=0$ only the quadratic replica coupling survives in the Hamiltonian.
At this point, the proof proceeds exactly like in \cite{quadratic}:
one  introduces an auxiliary Gaussian standard random variable $z$
with probability distribution
$$
d\mu(z)=e^{-z^2/2}\frac{dz}{\sqrt{2\pi}}
$$
and performs a simple rescaling, to write
\begin{eqnarray}
&&\frac1{2N}\ln \sum_{\{\sigma^1,\sigma^2\}}
e^{N {\lambda_0}\,q_{12}^2/2}=
\frac1{2N}\ln\sum_{\{\sigma^1,\sigma^2\}}
\int e^{\sqrt{\lambda_0 N}q_{12}z}d\mu(z)\\
\label{asdf}
&&=\ln 2+\frac1{2N}\ln\int \sqrt{\frac{N}{2\pi}}
\exp N\left(-\frac{y^2}2+\ln\cosh
\left(y\sqrt{\lambda_0}
\right)
\right).
\end{eqnarray}
For $\lambda_0=\lambda+2\alpha
E\tanh^2(\beta J)<1$, one can employ the inequality
\begin{equation}
2\ln\cosh x\le x^2
\end{equation}
to deduce, from Eqs. (\ref{integrando}) and (\ref{asdf}),
\begin{equation}
\label{boundsup}
\frac1{2N}E\ln Z_N^{(2)}(\beta,\alpha,\lambda;{\cal J})\le
\frac1N\ln E Z_N(\beta,\alpha;{\cal J})+\frac1{4N}\ln \frac1{1-\lambda_0},
\end{equation}
so that Eq. (\ref{sothat}) reduces to
\begin{eqnarray}
\nonumber
\frac \partial{\partial\alpha} \left(F_{RS}(\beta,\alpha;0)-\frac{
E\ln Z_N}N\right)&\le& \frac{2E \ln(1-\tanh^2(\beta J))^{-1}}{\lambda}\left(F_{RS}(\beta,\alpha;0)-\frac{
E\ln Z_N}N\right)\\\nonumber
&&+O(N^{-1}).
\end{eqnarray}
As in \cite{quadratic}, this implies
\begin{equation}
\label{stat0}
\left(F_{RS}(\beta,\alpha;0)-\frac{E\ln Z_N}N\right)=O(N^{-1}),\;\;\;\mbox{for}\;\;\;2\alpha
E\tanh^2(\beta J)<1,
\end{equation}
since
$$
\left.\left(F_{RS}(\beta,\alpha;0)-\frac{E\ln Z_N}N\right)\right|_{\alpha=0}=0.
$$
Statement (\ref{stat1}) then follows if one notices that, thanks to
(\ref{boundsup}), (\ref{stat0})
and to monotonicity of the free energy with respect to $\lambda$,
\begin{equation}
F_{RS}(\beta,\alpha;0)+O(N^{-1})=\frac1NE\ln Z_N\le \frac1{2N}
E\ln Z_N^{(2)}(\lambda)\le F_{RS}(\beta,\alpha;0)+O(N^{-1})
\end{equation}
in the region (\ref{triangular}).
Finally, statement (\ref{stat2}) follows from (\ref{stat1}) and from
convexity of $E\ln Z_N^{(2)}$ with respect to $\lambda$.
\hfill $\Box$


\section{Multi-overlap fluctuations in the annealed region}

In the previous section, we proved that the multi-overlap among any $2n$ configurations
$\sigma^{(a_1)},\ldots,\sigma^{(a_{2n})}$ is typically
small, in the annealed region. To study the infinite volume behavior of the multi-overlap fluctuations,
we define
$$
\eta_N^{a_1\ldots a_{2n}}=\sqrt N q_{a_1\ldots a_{2n}}\equiv\frac1{\sqrt N}\sum_{i=1}^N
\sigma^{(a_1)}_i\ldots\sigma^{(a_{2n})}_i.
$$
(Due to symmetry under permutation of the indices $a_i$, we will always assume them to be ordered as
$a_1<a_2<\ldots<a_{2n}$.) Then, like for the SK model at high temperature \cite{comets}, \cite{CLT}, \cite{T},
one can prove that the rescaled (multi)-overlaps
behave like independent centered Gaussian variables, in the infinite volume limit. Indeed, we prove the following
\begin{theorem}
\label{teoflutq}
In the annealed region (\ref{ann_reg}), the variables $\eta_N^{a_1\ldots a_{2n}}$
converge in distribution, as $N\to\infty$, to a centered Gaussian process $
\eta^{a_1\ldots a_{2n}}$ with covariances
\begin{eqnarray}
\label{covarianze1}
&&    \med{(\eta^{a_1\ldots a_{2n}})^2}=\frac1{1-2\alpha E\tanh^{2n}(\beta J)}\\
\label{covarianze2}
&&    \med{\eta^{a_1\ldots a_{2n}}\eta^{b_1\ldots b_{2n}}}=0\hspace{3cm}\mbox{if}\;\;\exists \;i:a_i\ne b_i\\
\label{covarianze3}
&& \med{\eta^{a_1\ldots a_{2n}}\eta^{a_1\ldots a_{2n'}}}=0\hspace{2.8cm}\mbox{if}\;\; n\ne n'.
\end{eqnarray}
\end{theorem}
{\bf Remark} Notice that, when the boundary of the annealed region (\ref{ann_reg}) is approached, only
the variance of $\eta^{a_1 a_2}$ diverges.

{\em Proof of Theorem \ref{teoflutq}}.
For simplicity, here we prove only that
\begin{equation}
\label{semplicita}
    \phi_N(u)\equiv\Med{e^{i\, u\, \eta_N^{12}}}\longrightarrow \exp\left(- \frac {u^2}{2(1-2\alpha E\tanh^2 (\beta J))}\right).
\end{equation}
The generalization (\ref{covarianze1}) to $n>1$ and the proof of the independence
(\ref{covarianze2})-(\ref{covarianze3}) of the limit random variables, though technically heavier,
present no additional conceptual difficulty. 

The proof is based on the cavity method \cite{MPV}
 (see \cite{guerracav}, \cite{talaRSB} and, in particular, \cite{talaksat}), which in essence consists in analyzing what happens when one removes one of the spins, thereby transforming the original system into one of size $N-1$.
As in \cite{CLT}, the idea is to write down a linear differential equation for $\phi_N(u)$, in the
thermodynamic limit. First of all, using symmetry among sites one can write
\begin{eqnarray}
\partial_u \phi_N(u)=i\Med{\eta^{12}_N e^{i\,u\,\eta^{12}_N}}=i\sqrt N \Med{\sigma^1_N
\sigma^2_N e^{i\,u\,\eta^{12}_N}}.
\end{eqnarray}
Notice that, thanks to Theorem \ref{teoaccopp} of the previous section,
\begin{equation}
    \left|\partial_u \phi_N(u)\right|\le \med{(\eta^{12}_N)^2}^{\frac12}\le C,
\end{equation}
uniformly in $N$. Then, defining
$$
u'=u\sqrt{1-1/N},
$$
one has
\begin{eqnarray}
\label{quasi}
\partial_u \phi_N(u)&=&i\sqrt N \Med{\sigma^1_N\sigma^2_N \exp\left(iu\sigma^1_N\sigma^2_N/\sqrt N+
i u'\eta^{12}_{N-1}\right)}\\
&=&-u \phi_N(u)+i\sqrt N\Med{\sigma^1_N\sigma^2_N e^{i u'\eta^{12}_{N-1}}}+o(1)
\end{eqnarray}
where the term $o(1)$, vanishing for $N\to\infty$, arises from the
expansion of $\exp( i u \sigma^1_N\sigma ^2_N/\sqrt N)$ around $u=0$ and from the replacement $u'\to u$.
Now consider the set
\begin{equation}
{\cal A}=\left\{{\cal J}:\nexists\, \mu:i_\mu=j_\mu=N\right\},
\end{equation}
where $i_\mu,j_\mu$ are the random site indices appearing in (\ref{H}).
Since the probability of ${\cal A}$ is very close to one,
$$
P({\cal A})=1-O(1/N),
$$
one can write
\begin{eqnarray}
\label{intermedio}
i\sqrt N\Med{\sigma^1_N\sigma^2_N e^{i u'\eta^{12}_{N-1}}}=
i\sqrt N\Med{\sigma^1_N\sigma^2_N e^{i u'\eta^{12}_{N-1}}1_{{\cal A}}}+o(1)
\end{eqnarray}
where $1_{\cal A}$ is the indicator function of the set ${\cal A}$. Next,
we  single out all terms $-J_\nu \sigma_{i_\nu}\sigma _N$ in the Hamiltonian (\ref{H}) involving the N-th spin
(the number of these terms
is a Poisson variable $\xi_{2\alpha}$ of mean value $2\alpha$) and we rewrite (\ref{intermedio}) as
\begin{eqnarray}
\label{AB}
i\sqrt NE\frac{\Omega'\left(e^{i\,u'\,\eta^{12}_{N-1}}Av\, \sigma^1_N \sigma^2_N \exp(\beta \sum_{\ell=1,2}\sigma^\ell_N
\sum_{\nu=1}^{\xi_{2\alpha}}J_\nu\sigma^\ell_{i_\nu})\right)}
{\Omega'\left(Av\, \exp(\beta \sum_{\ell=1,2}\sigma^\ell_N\sum_{\nu=1}^{\xi_{2\alpha}}J_\nu\sigma^\ell_{i_\nu})\right)}
\equiv i\sqrt NE\frac A B,
\end{eqnarray}
where $Av$ denotes average on the two-valued unbiased variables $\sigma^\ell_N=\pm1$ and $\Omega'(.)$ is the
Gibbs average for a system with $N-1$ spins and connectivity parameter $\alpha'=\alpha(1-1/(N-1))$\footnote{This is because
the average number of terms appearing in the modified Hamiltonian of the $N-1$ spin system is
$N\alpha-2\alpha\equiv\alpha'(N-1)$}.
Of course, since we are restricting to the set ${\cal A}$, the indices $i_\nu$ are i.i.d. random variables
uniformly distributed on $\{1,\ldots,N-1\}$. Now, we show that the denominator $B$ in (\ref{AB}) can be replaced
by the random variable
\begin{equation}
\tilde B=\prod_{\nu=1}^{\xi_{2\alpha}}\cosh^2(\beta J_\nu),
\end{equation}
by neglecting an error term which vanishes in the thermodynamic limit.
To this purpose we use the obvious identity
\begin{equation}
\label{identita}
E\frac AB=2E\frac A{\tilde B}-E\frac {AB}{\tilde B^2}+E\frac AB \left(\frac{\tilde B-B}{\tilde B}\right)^2,
\end{equation}
as it was done in \cite{talaRSB}.
As we will show below, the last term in the r.h.s. vanishes for $N\to\infty$.
The first term is easily computed. Indeed, recalling the mutual independence of the variables $J_\nu,
i_\nu$ and using the formula
$$
E a^{\xi_\lambda}=e^{-\lambda(1-a)},
$$
which holds for $a\ne0$ if $\xi_\lambda$ is a Poisson random variable of mean $\lambda$,
one finds
\begin{eqnarray}\nonumber
i\sqrt N E \frac A{\tilde B}&=& i\sqrt N
E\,\Omega'\left\{e^{i u' \eta^{12}_{N-1}}\sinh\left(2\alpha E\tanh^2(\beta J)
\frac{\eta^{12}_{N-1}}{\sqrt{N-1}}\right)\right\}.
\end{eqnarray}
Then,
expanding  the $\sinh(\ldots)$ at first order around zero and recalling that
$$
\sup_N \med{(\eta^{12}_N)^2}<\infty,
$$
one has
\begin{eqnarray}\label{pezzo1}
i\sqrt N E \frac A{\tilde B}&=& 2i\alpha E\tanh^2(\beta J)
E\,\Omega'\left\{e^{i u' \eta^{12}_{N-1}}\eta^{12}_{N-1}\right\}+o(1)\\
\nonumber &=&2\alpha E\tanh^2(\beta J)\partial_u
\phi_N(u)+o(1).
\end{eqnarray}
As for the second term in (\ref{identita}), one finds again
\begin{eqnarray}\label{pezzo2}
i\sqrt N E \frac {AB}{\tilde B^2}=2\alpha E\tanh^2(\beta J)\partial_u
\phi_N(u)+o(1).
\end{eqnarray}
Finally, we show that the last term can be neglected. First of all,
one has
$$
B\ge 1,
$$
as it follows from Jensen inequality,
interchanging the thermal average $\Omega'$ and the exponential in the definition of $B$. Therefore,
\begin{eqnarray}
\label{83}
\sqrt N \left|E\frac A B\left(\frac{\tilde B-B}{\tilde B}\right)^2
\right|\le \sqrt{N} Ee^{2\beta\xi_{2\alpha}}\left(1-\frac B{\tilde B}\right)^2.
\end{eqnarray}
The computation of (\ref{83}) proceeds in analogy with that of
$E A/\tilde B$. In this case, however, one finds that the dominant term in the Taylor expansion is of order
\begin{equation}
\label{pezzo3}
\frac1{\sqrt N}\med{(\eta^{12}_N)^2}=o(1).
\end{equation}
Therefore, recalling Eqs. (\ref{identita}), (\ref{pezzo1}), (\ref{pezzo2}), together with Eq. (\ref{quasi}), we find that $\phi_N(u)$ solves the linear differential equation
\begin{eqnarray}
\left(1-2\alpha E\tanh^2(\beta J)\right)\partial_u\phi_N(u)=-u\phi_N(u)+o(1)
\end{eqnarray}
which, together with the obvious initial condition
\begin{equation}
\phi_N(0)=1,
\end{equation}
implies the result (\ref{semplicita}).
\hfill  $\Box$

\section{Free energy fluctuations}

Is easy to realize that the Viana-Bray model resembles
locally a spin glass model on a tree, where the number of branches starting at
each node is a Poisson random variable of parameter $2\alpha$ and
the couplings associated to the branches are i.i.d. random variables $J_\mu$. The non-triviality
of the Viana-Bray model arises from the presence of
loops of length $O(\ln N)$ in the underlying graph.
For the model on the tree, the computation of the partition
function   for any disorder realization is elementary,
\begin{eqnarray}
Z_N^{tree}(\beta,\alpha;{\cal J})=2^N\prod_{\mu=1}^{\xi_{\alpha N}}\cosh(\beta
J_\mu),
\end{eqnarray}
so that
\begin{equation}
\frac1N E\ln Z_N^{tree}(\beta,\alpha;{\cal J})=\ln2+\alpha E\ln \cosh(\beta J).
\end{equation}
Theorem \ref{teoannealing} shows that, in the annealed region,
the Viana-Bray model behaves like its  tree-like counterpart, as far as
only the
infinite volume limit of the free energy density is concerned. However,
the difference
between the two models becomes evident  if one looks at the difference between the
respective free energies, on the scale $1/N$. Indeed, the following result
holds:
\begin{theorem}
\label{teoflutf}
Define the random variable
\begin{equation}
\hat f_N(\beta,\alpha;{\cal J})\equiv
\ln Z_N(\beta,\alpha;{\cal J})-(N\ln 2+\sum_{\mu=1}^{\xi_{\alpha N}}
\ln \cosh(\beta J_\mu)),
\end{equation}
where $J_1,\ldots, J_{\xi_{\alpha N}}$ are the same couplings which appear in the Hamiltonian
(\ref{H}).
In the annealed region (\ref{ann_reg}) $\hat f_N(\beta,\alpha;{\cal J})$
converges in distribution, as $N\to\infty$, to a non-Gaussian random variable
$\hat f$ with characteristic function
\begin{eqnarray}
\label{varlimite}
E\exp(i s \hat f)=\exp \left\{
-\frac12\sum_{n=1}^\infty is(is-1)\ldots (is-(2n-1))
\frac{\ln (1-2\alpha E\tanh^{2n}(\beta J))}{(2n)!}\right\}.
\end{eqnarray}
The variance of the limit random variable diverges when the boundary of the
annealed region is approached.
\end{theorem}
{\bf Remark} It is not difficult to check that, when the infinite
connectivity limit is performed as in Section 4, the limit random variable
becomes Gaussian (the terms of order higher than $s^2$ disappear in
the series) and one recovers the well known
result of Ref. \cite{alr} for the fluctuations of the SK free energy
at zero external field and $\beta'<1$.

{\em Proof of Theorem \ref{teoflutf}.}
The idea of the proof is to write down a linear differential
equation for the characteristic function
\begin{eqnarray}
\phi_N(\alpha,s)=E\,\exp(is \hat f_N).
\end{eqnarray}
Of course, for $\alpha=0$ both the Viana-Bray and the
tree model consist in an empty graph, so that
\begin{equation}
\phi_N(0,s)=1.
\end{equation}
As for the $\alpha$ derivative, the computation can be performed along the
lines of the computation of $\partial_\beta A_N(\beta,\alpha)$ in Section
3, with the result
\begin{eqnarray}
\frac{\partial\phi_N(\alpha,s)}{\partial\alpha}=
-N\phi_N(\alpha,s)+\frac1N\sum_{i,j=1}^N E \, e^{is\hat f_N}
\left(1+\tanh(\beta J)\omega_{\cal J}(\sigma_i\sigma_j)\right)
^{is}.
\end{eqnarray}
Since $|\tanh(\beta J)|<\tanh\beta<1$, one can expand the r.h.s. in an
absolutely convergent Taylor series, using the formula
$$
(1+x)^a=1+\sum_{n=1}^\infty \frac{a(a-1)\ldots(a-(n-1))}{n!} x^n
$$
and write
\begin{eqnarray}
\label{derivata}
\frac{\partial\phi_N(\alpha,s)}{\partial\alpha}=
\sum_{n=1}^\infty E\tanh^{2n}(\beta J)\frac{is(is-1)\ldots
(is-(2n-1))}{(2n)!}
E\,e^{is\hat f}\Omega_{\cal J}(N q_{1\ldots 2n}^2).
\end{eqnarray}
Notice that, thanks to Theorem \ref{teoaccopp},
$$
\med{Nq^2_{1\ldots 2n}}\le \med{Nq^2_{12}}\le\sup_N\med{Nq^2_{12}}<\infty
$$
and the derivative in (\ref{derivata}) can be bounded uniformly in $N$.
Next, we can replace $\Omega_{\cal J}(N q_{1\ldots 2n}^2)$ with
$\med{N q^2_{1\ldots 2n}}$. Indeed, thanks to Theorem \ref{teoflutq} of the previous
section,
\begin{eqnarray}
\Med{(\Omega_{\cal J}(N q_{1\ldots 2n}^2)-\med{N q^2_{1\ldots 2n}})^2}=
\med{(\eta^{1\ldots 2n}_N)^2(\eta^{2n+1\ldots 4n}_N)^2}-\med{(\eta^{1\ldots 2n}_N)^2}^2=
o(1).
\end{eqnarray}
Therefore, denoting by $\phi$ the infinite volume limit of $\phi_N$, one has
\begin{eqnarray}
\frac{\partial\phi(\alpha,s)}{\partial\alpha}=
\sum_{n=1}^\infty \frac{is(is-1)\ldots(is-(2n-1))}{(2n)!}
\frac{E\tanh^{2n}(\beta J)}{1-2\alpha E\tanh^{2n}(\beta J)}
\phi(\alpha,s),
\end{eqnarray}
from which the statement of the theorem follows after integration with
respect to $\alpha$.\hfill $\Box$

\section{Outlook and conclusions}

In this paper, we have provided a complete picture of the high temperature or low connectivity phase of the Viana-Bray
model, where annealing holds. Breaking of annealing is forecasted by the divergence of fluctuations of the free energy
density
(on the scale $1/N$) and of the two-replica overlap (on the scale $1/\sqrt N$). On the other hand, the fluctuations
of the multi-overlap among $2n\ge4$ configurations show no singularity when the boundary of the annealed region is
approached.

The high temperature phase of the diluted $p$-spin model with $p>2$ can be studied with the same techniques, but in this case one does not control the whole expected annealed region.
On the other hand, the methods we presented here do not extend directly to the study of the replica symmetric region of the K-sat model, or of the 
diluted mean field model in presence of a magnetic field. In
this case annealing does not hold, even at high temperature,
and the random variable $g$ which realizes the infimum of the replica symmetric functional 
is not trivial, as it is well known (see for instance \cite{mz}). We plan to report on this subject in a future paper.

\appendix

\section*{Appendix}

\section{Self-averaging of free energy and ground state energy densities}

In this section we prove an upper bound, exponentially small in $N$ and
independent of $\beta$, for
the fluctuations of the disorder dependent free energy density of the Viana-Bray model.
Independence of $\beta$ implies that the bound holds also for the
fluctuations of the ground state energy density.
Similar results have been known for a long time in the case of
fully connected mean field spin glass models (for instance, see
\cite{talagcorso} and references therein) and
for some random optimization problems \cite{talaconc}, \cite{rhee}.

\begin{theorem}
\label{teoconcentr}
For any value of $\beta,\alpha$ and $N$, one has
\begin{equation}
\label{stima2}
P\left(\left| \frac1{N\beta}\ln Z_N
-\frac1{N\beta}E\ln Z_N\right|\ge u\right)
\le 2e^{ N\left(
u-\alpha(1+\frac u{\alpha})\ln(1+\frac u{\alpha})\right)}.
\end{equation}
\end{theorem}
{\bf Remarks}
The theorem can be immediately extended
to the more general class of
diluted spin glass models considered in \cite{franz-leone}.
In particular, for the diluted $p$-spin model \cite{ricci}  with $p\ge 3$
the above result holds without
any modification, while for the K-sat model
one has to replace  (\ref{stima2}) by
\begin{eqnarray}
\label{ksat1}
&&P\left(\frac{\ln Z_N}{N\beta}-\frac{E\ln Z_N}{N\beta}\le -u\right)\le e^{ N\left(
u-\alpha(1+\frac u{\alpha})\ln(1+\frac u{\alpha})\right)}\;\;\;\;\;\;u>0\\
\label{ksat2}
&&P\left(\frac{\ln Z_N}{N\beta}-\frac{E\ln Z_N}{N\beta}\ge u\right)\le
e^{ N\left(-
u-\alpha(1-\frac u{\alpha})\ln(1-\frac u{\alpha})\right)}\;\;\;0<u<\alpha\\
&&P\left(\frac{\ln Z_N}{N\beta}-\frac{E\ln Z_N}{N\beta}\ge u\right)=0\hspace{3.9cm}u\ge\alpha.
\end{eqnarray}
(The latter is a simple consequence
of the fact that, for the K-sat, $1/N \ln Z_N\le \ln2$ for any
disorder realization, and that $1/N E\ln Z_N\ge \ln 2-\alpha\beta$,
as it is easily verified from the definition of the model.)
In particular, Eqs. (\ref{ksat1})-(\ref{ksat2}) allow to recover the bound given in \cite{broder}
for the fluctuations of the minimal fraction of unsatisfied clauses in
the K-sat problem.

{\em Proof of Theorem \ref{teoconcentr}}. We sketch just the main steps in the proof, since it is very similar
in spirit to that given for fully connected models in \cite{talagcorso} \cite{limterm2},
the main difference being that the role of
Gaussian integration by parts is replaced
here by  the properties (\ref{prop1}), (\ref{prop2}) of
Poisson random variables.

Introduce the interpolating parameter $0\le t\le1$ and define, for $s\in{\mathbb R}$,
\begin{equation}
\varphi_N(t)=\ln E_1\exp\left\{s E_2\ln Z_N(t)\right\},
\end{equation}
where
\begin{eqnarray}
Z_N(t)=\sum_{\{\sigma\}}\exp\beta\left(\sum_{\mu=1}^{\xi^1_{2\alpha Nt}}
J^1_\mu\sigma_{i^1_\mu}\sigma_{j^1_\mu}+
\sum_{\nu=1}^{\xi^2_{2\alpha N(1-t)}}
J^2_\nu\sigma_{i^2_\nu}\sigma_{j^2_\nu}\right).
\end{eqnarray}
Here, all variables with upper index $1$ are independent from those with index
$2$, and $E_\ell$ denotes the average
$$
E_\ell(.)=E_{\xi^\ell}E_{\{J^\ell_\mu\}}E_{\{i^\ell_\mu\}}E_{\{j^\ell_\mu\}}(.),\hspace{1cm}\ell=1,2.
$$
The motivation for the introduction of $\varphi_N(t)$ is the identity
\begin{eqnarray}
\label{identity}
\exp\{\varphi_N(1)-\varphi_N(0)\}=E\,\exp \left\{s\left(\ln Z_N-E \ln Z_N\right)\right\}.
\end{eqnarray}
Since we want to bound the r.h.s. of (\ref{identity}), we compute the
$t$ derivative of $\varphi_N(t)$. After some straightforward computations, one
finds
\begin{eqnarray}
\nonumber
\varphi'_N(t)=\alpha\frac{\sum_{i,j=1}^N}N
\frac{E_1\left\{e^{s E_2\ln Z_N(t)}
E_J\left(e^{s E_2\ln \omega(e^{\beta J \sigma_i\sigma_j})}-1-
s E_2\ln \omega (e^{\beta J \sigma_i\sigma_j})\right)\right\}}
{E_1\exp\left\{s E_2\ln Z_N(t)\right\}}
\end{eqnarray}
and, thanks to the trivial bounds
$$
-\beta \le E_2\ln \omega(e^{\beta J\sigma_i\sigma_j})\le \beta
$$
one has
\begin{equation}
\label{putting}
|\varphi'_N(t)|\le \alpha N(e^{|s|\beta}-1-|s|\beta ).
\end{equation}
Putting together Eqs. (\ref{putting}) and (\ref{identity}),
employing Tchebyshev's
 inequality and optimizing on $s$, one finally
finds the statement of the theorem.
\hfill $\Box$



\vspace{.5cm}
{\bf Acknowledgments}

F.L.T. wishes to thank in particular Erwin Bolthausen for useful conversations and for his kind
invitation to the Institute of Mathematics of the University of Zurich,
where a large part of this work was done.

This work was supported in part by Swiss Science Foundation,
by MIUR
(Italian Minister of Instruction, University and Research),
and by INFN (Italian National Institute for Nuclear Physics).


\begin{thebibliography}{199}

\bibitem{MPV} M. M\'ezard, G. Parisi and M. A. Virasoro, {\sl Spin glass theory
and beyond}, World Scientific, Singapore (1987).

\bibitem{1RSB} M. M\'ezard, G. Parisi, R. Zecchina,
{\em Analytic  and Algorithmic Solution of Random Satisfiability Problems},
Science {\bf 297},  812-815 (2002).

\bibitem{franz-leone} S. Franz, M. Leone,
{\em Replica bounds for optimization problems and diluted spin systems},
to appear on J. Stat. Phys., {\tt cond-mat/0208280}.

\bibitem{talaksat} M. Talagrand, {\em
The high temperature case of the K-sat problem}, Probab. Theory Rel. Fields
{\bf 119}, 187-212 (2001).

\bibitem{limterm} F. Guerra, F. L. Toninelli, {\sl
The Thermodynamic Limit in Mean Field Spin Glass Models},
Commun. Math. Phys. {\bf 230:1}, 71-79 (2002).

\bibitem{broken}  F. Guerra, {\em
Broken Replica Symmetry Bounds in the Mean Field Spin Glass Model},
Commun. Math. Phys. {\bf 233:1}, 1-12 (2003).

\bibitem{viana} L. Viana, A. J. Bray, {\em Phase diagrams for dilute
spin-glasses}, J. Phys. C {\bf 18}, 3037-3051 (1985).

\bibitem{kanter} I. Kanter, H, Sompolinsky, {\em Mean-Field Theory of Spin-Glasses with Finite
Coordination Number}, Phys. Rev. Lett {\bf 58}, 164 (1987).

\bibitem{sk} D. Sherrington, S. Kirkpatrick, {\sl Solvable model of
a spin-glass}, Phys. Rev. Lett. {\bf 35}, 1792-1796 (1975).

\bibitem{quadratic} F. Guerra, F. L. Toninelli,
{\sl Quadratic replica coupling for
the Sherrington-Kirkpatrick mean field spin glass model}, J. Math. Phys.
{\bf 43}, 3704-3716 (2002).

\bibitem{CLT} F. Guerra, F.~L.~Toninelli, {\em Central limit theorem for
fluctuations in the high temperature region of the
Sherrington-Kirkpatrick spin glass model}, J. Math. Phys. {\bf 43},
  6224-6237 (2002).

\bibitem{xor-sat} N. Creignou, H. Daude, {\em Satisfiability threshold
for random XOR-CNF formulas}, Discr. Appl. Math. {\bf 96-97}, 41-53 (1999).

\bibitem{talagcorso} M. Talagrand, {\sl Mean field models for spin glasses:
a first course}, to appear in the Proceedings of the 2000 Saint
Flour Summer School in probability.

\bibitem{comets} F. Comets, J. Neveu, {\em The Sherrington-Kirkpatrick model of spin glasses and
stochastic calculus: the high temperature case}, Commun. Math. Phys. {\bf 166}, 549-564 (1995).

\bibitem{T} M. Talagrand,
{\sl Spin glasses: a challenge for mathematicians. Mean field models and
cavity method}, Springer Verlag, to appear.

\bibitem{guerracav} F. Guerra,
{\em The cavity method in the mean field spin glass model. Functional representation of the thermodynamic variables},
in: Advances in Dynamical Systems and Quantum Physics, S. Albeverio, {\em et al.} eds., World Scientific, Singapore (1995). 

\bibitem{talaRSB} M. Talagrand, {\em Exponential inequalities and Replica Symmetry Breaking for the Sherrington-Kirkpatrick Model}, Ann. Probab. {\bf 28}, 1018-1062 (2000).


\bibitem{alr} M. Aizenman, J. Lebowitz and D. Ruelle, {\sl Some rigorous
results on the Sherrington-Kirkpatrick spin glass model},
Commun. Math. Phys. {\bf 112}, 3-20 (1987).

\bibitem{mz} R. Monasson, R. Zecchina, {\em Statistical mechanics of the random K-satisfiability model},
Phys. Rev. E {\bf 56}, 1357 (1997).

\bibitem{talaconc} M. Talagrand, {\em
Concentration of measure and isoperimetric inequalities in product spaces},
Publ. Math. I.H.E.S. {\bf 81}, 73-205 (1995).

\bibitem{rhee} W. Rhee, M. Talagrand, {\em
Martingale inequalities and NP-complete problems}, Math. Oper. Res. {\bf
 12}, 177-181 (1987).

\bibitem{ricci} F. Ricci-Tersenghi, M. Weigt,  R. Zecchina, {\em Simplest random K-satisfiability problem},
Phys. Rev. E {\bf 63}, 026702 (2001).

\bibitem{broder} A. Z. Broder, A. M. Frieze, E. Upfal, {\em
On the Satisfiability and Maximum Satisfiability of Random 3-CNF
Formulas}, in {\em Proceedings of the 4th Annual ACM-SIAM Symposiym on Discrete
Algorithms} (Association for Computing Machinery, New York, 1993), p. 322-330.

\bibitem{limterm2} F. Guerra, F. L. Toninelli, {\sl
The infinite volume limit in generalized mean field disordered models},
to appear on Markov Proc. Rel. Fields,  {\tt cond-mat/0208579}.



\end{thebibliography}
\end{document}